\documentclass[preprint,amsmath,amssymb,a4]{revtex4-1}   
\usepackage{graphicx}
\usepackage{bm}
\usepackage{color}
\usepackage{ulem}
 
 \usepackage{epsfig}
\usepackage{chngcntr}
\counterwithout{figure}{section}
\usepackage{rotating}
\usepackage{physics}
\usepackage{multirow}
\usepackage{float}
\usepackage{caption}

\begin{document}

\title{Solution of the random field $XY$ magnet on a fully connected graph}

\author{Sumedha,}
\email{sumedha@niser.ac.in}
\address{School of Physical Sciences, National Institute of Science Education and Research, Bhubaneswar, P.O. Jatni, Khurda, Odisha, India 752050; Homi Bhabha National Institute, Training School Complex, Anushakti Nagar, Mumbai, India 400094}
\author{Mustansir Barma}
\email{barma@tifrh.res.in}
\address{TIFR Centre for Interdisciplinary Sciences, Tata Institute of Fundamental Research, Gopanpally, Hyderabad, India  500046}

\begin{abstract}

We use large deviation theory to obtain the free energy of the XY model on a fully connected graph on each site of which there is a randomly oriented field of magnitude $h$. The phase diagram is obtained for two symmetric distributions of the random orientations: (a) a uniform distribution and (b) a distribution with cubic symmetry. In both cases, the disorder-averaged ordered state reflects the symmetry of the underlying distribution. The phase boundary has a multicritical point which separates a locus of continuous transitions (for small values of $h$) from a locus of first order transitions (for large $h$). The free energy is a function of a single variable in case (a) and a function of two variables in case (b), leading to different characters of the multicritical points in the two cases. We find that the locus of continuous transitions is given by the same equation for a family of quadriperiodic  distributions, which includes the distributions (a) and (b). However, the location of the multicritical point and the nature of ordered state depend on the form of the distribution. The disorder-averaged ground state energy is found exactly, and the specific heat is shown to approach a constant as temperature approaches zero.

\end{abstract}
\maketitle

\section{Introduction}

Random disorder in the field conjugate to the order parameter is known to have important effects in a number of contexts. A random field model was first introduced by Larkin \cite{larkin} to model vortex lattices in superconductors. Later it was used to model disordered antiferromagnets in a uniform field \cite{fishman,belanger}, binary fluids in random porous media \cite{vink}, phase transitions in random alloys \cite{maher}, charge density waves with impurity pinning \cite{lee} and 
social interactions via network models \cite{michard}. For such systems, Imry and Ma argued that arbitrarily weak random field disorder would destroy an ordered phase for all dimensions less than two (four) for discrete (vector) spins \cite{imryma}. The case of discrete Ising spins, namely the random field Ising model (RFIM) has been particulary well studied \cite{schneider,aharony1,fytas1,fytas}. Vector spins behave differently from discrete spins both in the absence and presence of disorder, due to low lying modes and topological features  \cite{proctor}. Also, random field models with vector spins have been argued to exhibit a spin glass state \cite{cardy,doussal}.

Disordered vector-spin models on networks with long-range connectivity are of current interest. The case of random graphs in which the Hausdorff dimension of the lattice is infinite has several applications.  For instance, XY spin models on random graphs have been applied to the study of structural phase transitions in networks \cite{mendes, kwak,yang}. Also the disordered ferromagnetic XY model is close to the Kuramoto model with quenched random frequencies which describes the phenomenon of synchronization \cite{collet}. Recently the XY model has also been used to study the Markov Random field models \cite{wada} and neural networks \cite{stroev}. 

In this paper we use large deviation theory (LDT) to find the exact free energy of the random field XY (RFXY) model on a fully connected graph in the thermodynamic limit for various distributions of the random field. Earlier, random field vector models were studied using methods based on mean field theory \cite{aharony1,saxena}, replica methods \cite{cardy,doussal}, variational principles \cite{garel}, effective field theory \cite{eft}, renormalization group \cite{aharony1,perret} and belief propagation \cite{lupo}.  As discussed below, our study resolves certain discrepancies in the nature of the reported phase boundaries, and also addresses thermodynamic properties at and close to $T=0$.

In a recent study, Lupo et al studied the RFXY model with uniformly distributed orientation of the random field of magnitude $h$ on a regular random graph with finite connectivity using the belief propagation method \cite{lupo}. They reported a replica symmetry broken phase at low temperatures, associated with spin glass order. They also considered the model on a fully connected graph (the SK limit) and found a continuous transition, along with a re-entrant phase boundary in the $T-h$ plane. On the other hand, in earlier work,  Aharony had studied the random field Heisenberg model on a fully connected graph \cite{aharony1}. It was argued that for a symmetric random field distribution with a minimum at zero strength of the random field, the transition would become first order at a sufficiently low temperature $T$. Consistent with this, Aharony found a tricritical point separating second order and first order transitions for a Heisenberg model. The LDT results reported in \cite{sumedhaetal} and discussed below, yield a phase diagram which includes first order transitions, agreeing with \cite{aharony1}. Further, we find that the locus of transitions does not exhibit re-entrance, unlike in \cite{saxena, eft,lupo}. Some of the results of \cite{lupo} have been corrected in \cite{lupoc}.

We obtain the phase diagram for uniform and cubic distributions of the orientation of the random field. In both cases, the disorder-averaged state at low temperatures reflects the symmetry of the distribution. Along the phase boundary separating the disordered and ordered phases we show that there is a multicritical point (MCP) which separates continuous transitions (for small values of $h$) from first order transitions (for large $h$). Interestingly, we find that the locus of continuous transitions in the $T-h$ plane is given by the same equation for a family of distributions which includes the uniform and cubic cases. However, the nature and location of the MCP does depend on the distribution of random fields as does the locus of first order transitions. We also find analytic forms for the exact ground state energy for both uniform and cubic distributions, and demonstrate a first order jump of the magnetization at $T=0$.


The calculation described here is the first application of LDT to a problem with disorder for continuous spins, though the rate function for the fully connected graph was derived recently in the absence of disorder \cite{kirkpatrick}. Recently LDT \cite{touchette} was used to perform disorder averaging for discrete-spin random-field quenched disorder problems on a fully connected graph \cite{lowe,sumedhasingh,sumedhajana,kistler}, but the current study differs in important ways from these problems, especially at low temperatures where the vector character of the spins affects thermodynamic properties significantly.

The plan of the paper is as follows :  In Section \ref{sec1}, we obtain the expression for the rate function, for arbitrary distribution of the random field orientation using LDT. In Section \ref{sec2} we use the rate function to obtain the full diagram for uniform and cubic distributions in the $T-h$ plane. We also obtain the equation for the locus of continuous transitions for a family of distributions. In Section \ref{sec3} we study the behaviour of various thermodynamic quantities near different types of phase transitions for the cubic distribution. In Section \ref{sec4} we study the zero and low temperature properties of the model and conclude in Section \ref{sec5}.

\section{Random field XY model via Large deviation theory}
\label{sec1}
The Hamiltonian of the RFXY model on a fully connected graph can be written as:

\begin{equation}
H=-\frac{1}{2 N} (\sum_{i=1}^N \vec{s_i})^2 -h \sum_{i=1}^N (\vec{n_i}. \vec{s_i})
\label{Hxy}
\end{equation}
where $\vec{s_i}= \cos \theta_i \hat{i}+ \sin \theta_i \hat{j}$ is an $m=2$ vector spin on a unit circle. Each pair of spins is coupled through an energy $(1/N) s_i. s_j$. The angle $\theta_i$ is a random variable that lies in the interval $[0,2 \pi]$. The constant $h$ represents the strength of the random field and $\vec{n_i} = \cos \alpha_i \hat{i}+\sin \alpha_i \hat{j}$ is a unit vector in its direction. Each $\alpha_i$ is an i.i.d chosen from a  distribution, $p(\alpha)$. 

Note that $H$ maybe rewritten as:
\begin{equation}
H=-\frac{1}{2 N} (\sum_{i=1}^N {\cos \theta_i {\hat{i}} + \sin \theta_i {\hat{j}}})^2 -h\sum_{i=1}^N \cos(\theta_i-\alpha_i)
\label{Hrf}
\end{equation}
and that in any configuration $C_N$ specified by the set of spin orientations ${\theta_i}$, the magnetizations along $x$ and $y$ directions are given by $x_1=\sum_{i=1}^N \cos \theta_i/N$ and $x_2= \sum_{i=1}^N \sin \theta_i/N$. 
 
 Below we obtain the properties of the system for arbitrary $p(\alpha)$ using large deviation theory.
\subsection*{Calculation of the free energy}
The probability $P_{H,\beta}$ of the occurence of a configuration $C_N$ is proportional to $\exp(-\beta H(C_N))$, where $\beta =1/T$. We assume that the random variables $(\sum_{i=1}^N \cos \theta_i,\sum_{i=1}^N \sin \theta_i)$ satisfy  the Large Deviation Principle (LDP) with respect to $P_{H,\beta}$, with a rate function $I(x_1,x_2)$. To leading order 
\begin{equation}
P_{H,\beta}(C_N:x_1,x_2) \sim \exp(-N I(x_1,x_2))
\label{ldp}
\end{equation}
In order to calculate $I(x_1,x_2)$, we first calculate the rate function $R(x_1,x_2)$ corresponding to the non-interacting part of the Hamiltonian $H_{ni}= -h\sum_{i=1}^N \cos(\theta_i-\alpha_i)$, using the G{\"a}rtner Ellis theorem \cite{touchette}.  The tilted large deviation principle \cite{hollander} then relates the two rate functions via the relation, $I(x_1,x_2)= -\beta (x_1^2+x_2^2)/2 +R(x_1,x_2)$ up to a constant term independent of $x_1,x_2$. 

Let us first calculate the rate function $R(x_1,x_2)$. The G{\"a}rtner Ellis theorem states that the rate function for the probability distribution of a random variable is given by the Legendre-Frenchel transform of the corresponding scaled cumulant generating function ($\Lambda(y_1,y_2)$):
\begin{equation}
R(x_1,x_2) =  sup_{y_1,y_2}  \{x_1 y_1 +x_2 y_2 -\Lambda(y_1,y_2)\}
\label{rfR}
\end{equation}
where $\Lambda(y_1,y_2)= \lim_{N \rightarrow \infty} \Lambda_N(y_1,y_2)/N$. The function $\Lambda_N(y_1,y_2)$ is the log cumulant generating function of the random variables $(\sum_{i=1}^N \cos \theta_i,\sum_{i=1}^N \sin \theta_i)$ for the probability distribution $P_{H_{ni},\beta}$. 
\begin{equation}
\Lambda_N(y_1,y_2) = log \left\langle \exp( y_1 \sum_{i=1}^N \cos \theta_i + y_2 \sum_{i=1}^N sin \theta_i) \right\rangle_Q
\end{equation}
Here $\langle ... \rangle_Q$ represents the expectation value w.r.t. 
the probability distribution $Q \propto e^{-\beta H_{ni}}$, which is a product measure over the probability distributions $Q_i$ for the non-interacting spins. Since $Q_i \propto exp(\beta h \cos (\theta-\alpha_i))$, we obtain
\begin{align}
\Lambda(y_1,y_2) &= \lim _{N \rightarrow \infty} \frac{\Lambda_N}{N} = \lim_{N \rightarrow \infty} \frac{1}{N} \sum_{i=1}^N \log S_i
\label{lambdainf}
\end{align}
where 
\begin{equation}
S_i=\frac{1}{\tilde{N}} \int_0^{2 \pi} d \theta \exp(\beta h \cos (\theta-\alpha_i) +y_1 cos \theta +y_2 sin \theta)
\label{S}
\end{equation}
Here $\tilde{N} = \int d \theta \exp(\beta h \cos (\theta-\alpha))$ is the normalisation.

Since $\alpha_i$ are i.i.d's chosen from a distribution $p(\alpha)$, the law of large numbers implies that as $N \rightarrow \infty$, Eq. \ref{lambdainf} becomes
\begin{equation}
\Lambda(y_1,y_2) =\int_{0}^{2 \pi} d \alpha~p(\alpha) \log S
\end{equation}

Let $(y_1^*,y_2^*)$ extremise the r.h.s of Eq. \ref{rfR}. Both $y_1^*$ and $y_2^*$ are functions of $x_1$ and $x_2$, given by the solutions of the equations:
\begin{align}
x_{1,2} &= \frac{\partial \Lambda(y_1,y_2)}{\partial y_{1,2}}
\end{align}
The rate function $I(x_1,x_2)$ can then be written as
\begin{align}
I(x_1,x_2) = g(x_1,x_2)- inf_{x_1,x_2}g(x_1,x_2)
\label{fullI}
\end{align}
where
\begin{equation}
g(x_1,x_2) = x_1 y_1^* +x_2 y_2^* -\Lambda(y
_1^*,y_2^*)-\frac{\beta (x_1^2+x_2^2)}{2}
\end{equation}
The free energy of the system is equal to $\frac{1}{\beta} inf_{x_1,x_2} I(x_1,x_2)$ \cite{ldbook}.

In the thermodynamic limit the probability $P_{H,\beta}(C_N:x_1,x_2)$  in  Eq. \ref{ldp} is dominated by minimum of $I(x_1,x_2)$, 
where $\frac{\partial I}{\partial x_1}=0$ and $\frac{\partial I}{\partial x_2}=0$, which yields $y_1^* = \beta x_1$ and $y_2^*=\beta x_2$. On substituting in Eq. \ref{fullI} we get a different function with the same extremal points, given by
\begin{eqnarray}
\label{grf}
I(x_1,x_2)&=&\frac{\beta}{2} (x_1^2+x_2^2)+\log I_0(\beta h)-\\\nonumber 
&&\int_0^{2 \pi} d \alpha p(\alpha) \log I_0(\beta \sqrt{h^2+(x_1^2+x_2^2)+2 h (x_1 \cos \alpha+x_2 \sin \alpha)})
\end{eqnarray}
here $I_0$ represents the zeroth modified Bessel function of the first kind. We have dropped the term $inf_{x_1,x_2}g(x_1,x_2)$ from the expression, as it is a constant which  displaces the entire functional and plays no role in determining the extremal points of the rate function. 

The expression for the rate function above is similar to the form of free energy obtained using mean field theory \cite{aharony1}. The function in Eq. \ref{grf} matches the function in Eq. \ref{ldp} at the extremum points for $x_1$ and $x_2$ lying in the interval $[-1,1]$. Hence both functions give rise to the same thermodynamic behaviour in the limit of infinite $N$. 

\section{Phase Diagram}
\label{sec2}
In this section, we obtain the phase diagram for two symmetric  distributions, namely  (a) uniform along the circle, and (b) cubic, with the field along $\pm \hat{i}$ or $\pm \hat{j}$. We then show that the locus of continuous transitions is given by the same equation in the generalized family of quadriperiodic distributions satisfying $p(\alpha)=p(\alpha+\pi/2)$.

\subsection{Uniform Distribution}
\label{suf}

Consider the case where the angle $\alpha$ is chosen uniformly from the interval $[0,2 \pi]$, i.e $p(\alpha)=1/(2 \pi)$. In this case though we cannot perform the integral in Eq. \ref{ldp} exactly, we may expand the integrand in a power series in $x_1$ and $x_2$ term by term and then integrate. The rotational symmetry of the distribution then implies that the rate function 
$I(x_1,x_2)$ is a function only of $r=\sqrt{x_1^2+x_2^2}$.


Expanding to $6^{th}$ order, we obtain
\begin{equation}
L(r) =c_2 r^2+\frac{c_4}{h^4} r^4+\frac{c_6}{h^6} r^6
\label{lc6}
\end{equation}
where the coefficients $c_4$ and $c_6$ are functions of $a=\beta h $ alone. We find:
\begin{eqnarray}
c_2&=& \frac{\beta}{2} \left(1-\frac{\beta}{2}+\frac{\beta I_1^2}{2 I_0^2}\right)
\label{c2u}
\end{eqnarray}
\begin{eqnarray}
c_4 &=& \frac{1}{32 I_0^4}(a^4 I_0^4- 2 a^3 I_0^3 I_1+2 a^2(1-2 a^2) I_0^2 I_1^2+4 a^3 I_0 I_1^3+3 a^4 I_1^4)
\label{lxyrfu}
\end{eqnarray}
where $I_0$ and $I_1$ are zeroth and first modified Bessel 
functions of the first kind respectively. Their argument 
$a = \beta h$ is not displayed explicitly.

\begin{figure}
\centering
\includegraphics[scale=0.7]{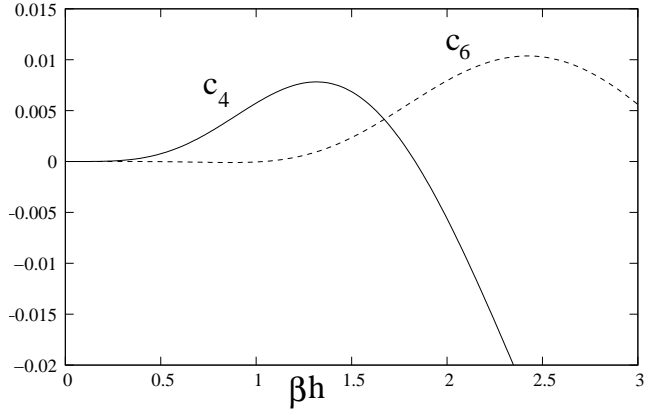}
\caption{The solid and dotted lines depict the coefficients of $r^4/ h^4$ and $r^6/ h^6$ terms in Eq. \ref{lc6}. The coefficient $c_4$ becomes negative at $\beta h=1.833$, at which point $c_6$ is positive. $c_6$ is slightly negative for $\beta h < 1.038$, and then again for $\beta h > 3.318$, at which point $c_8$ is positive.}
\label{c4}
\end{figure}

\begin{figure}
     \centering
         \includegraphics[width=0.45\textwidth]{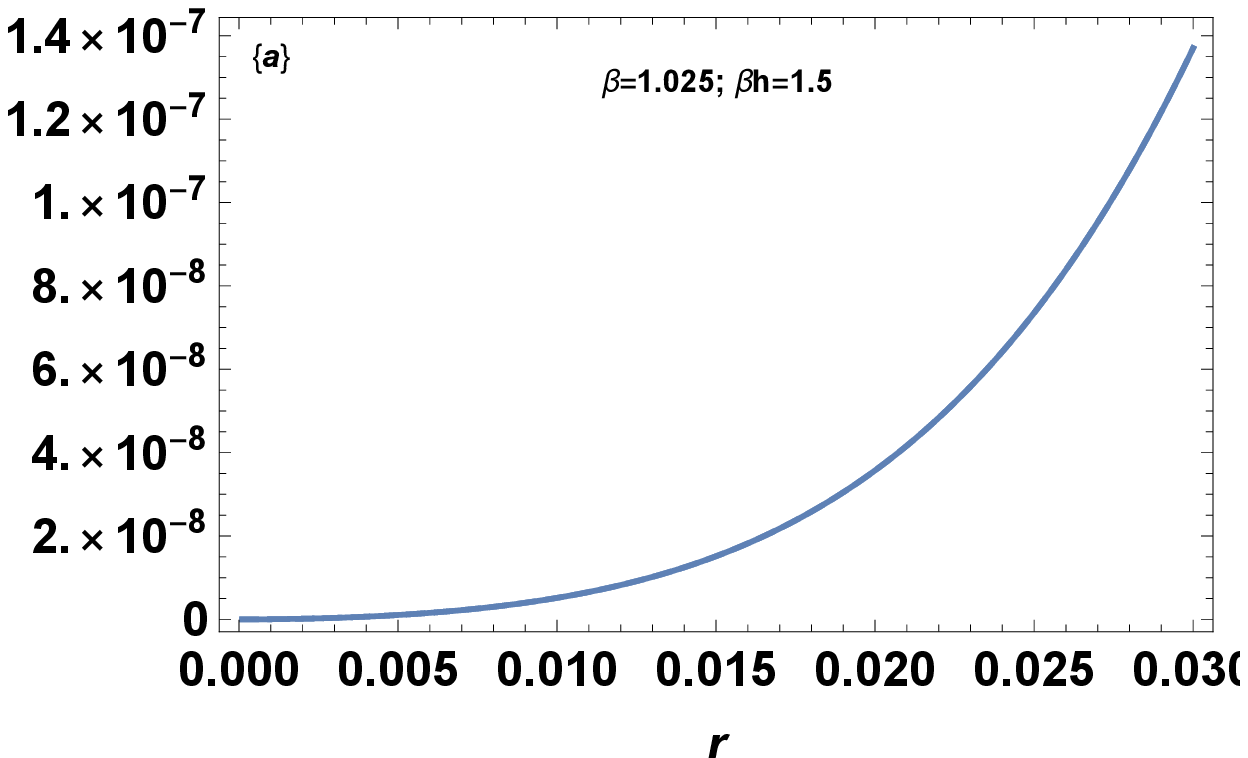}
     \hfill
         \includegraphics[width=0.45\textwidth]{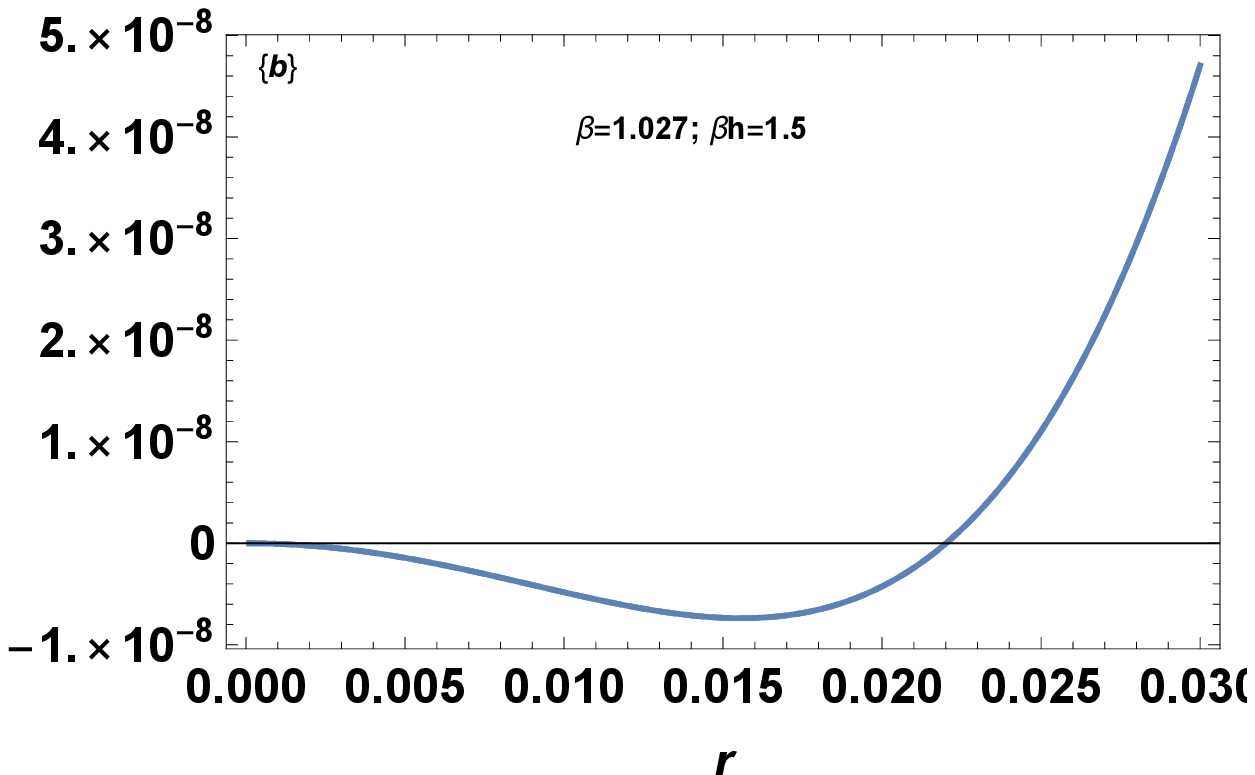}
     \hfill
         \includegraphics[width=0.45\textwidth]{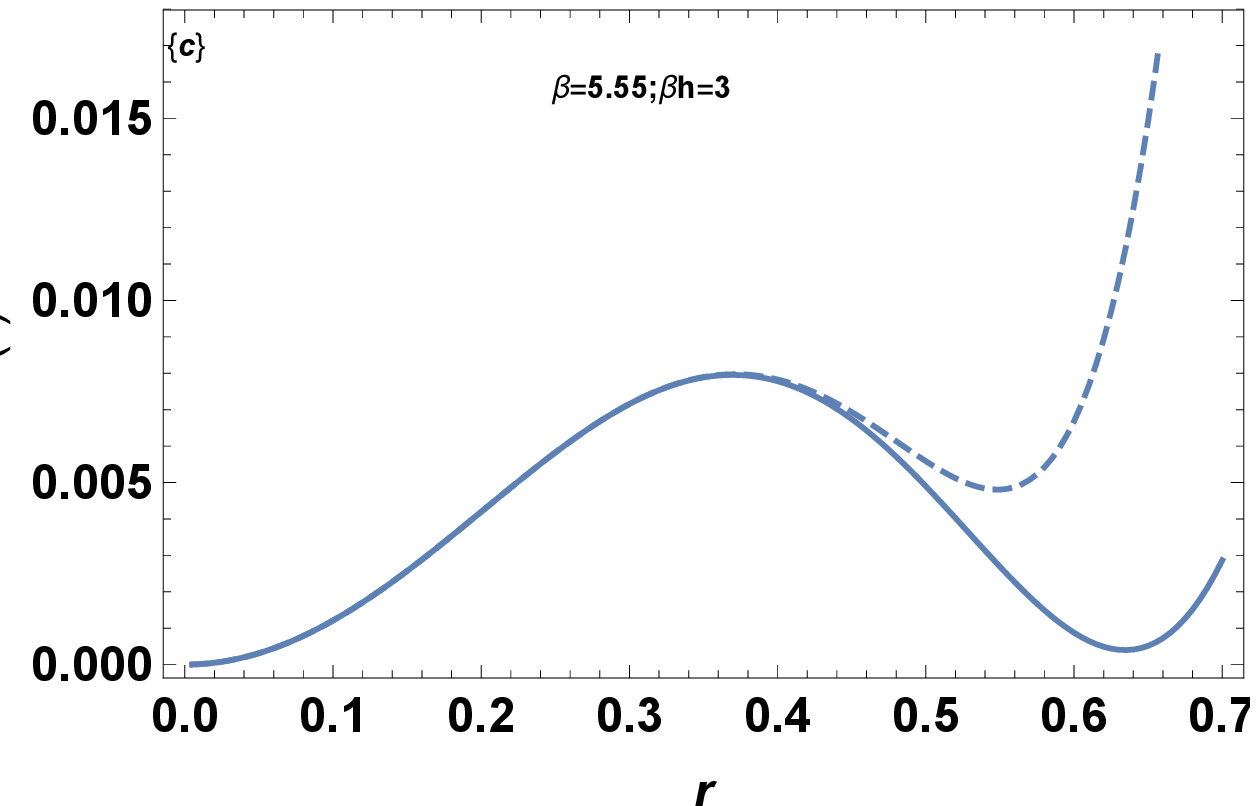}
     \hfill
         \includegraphics[width=0.45\textwidth]{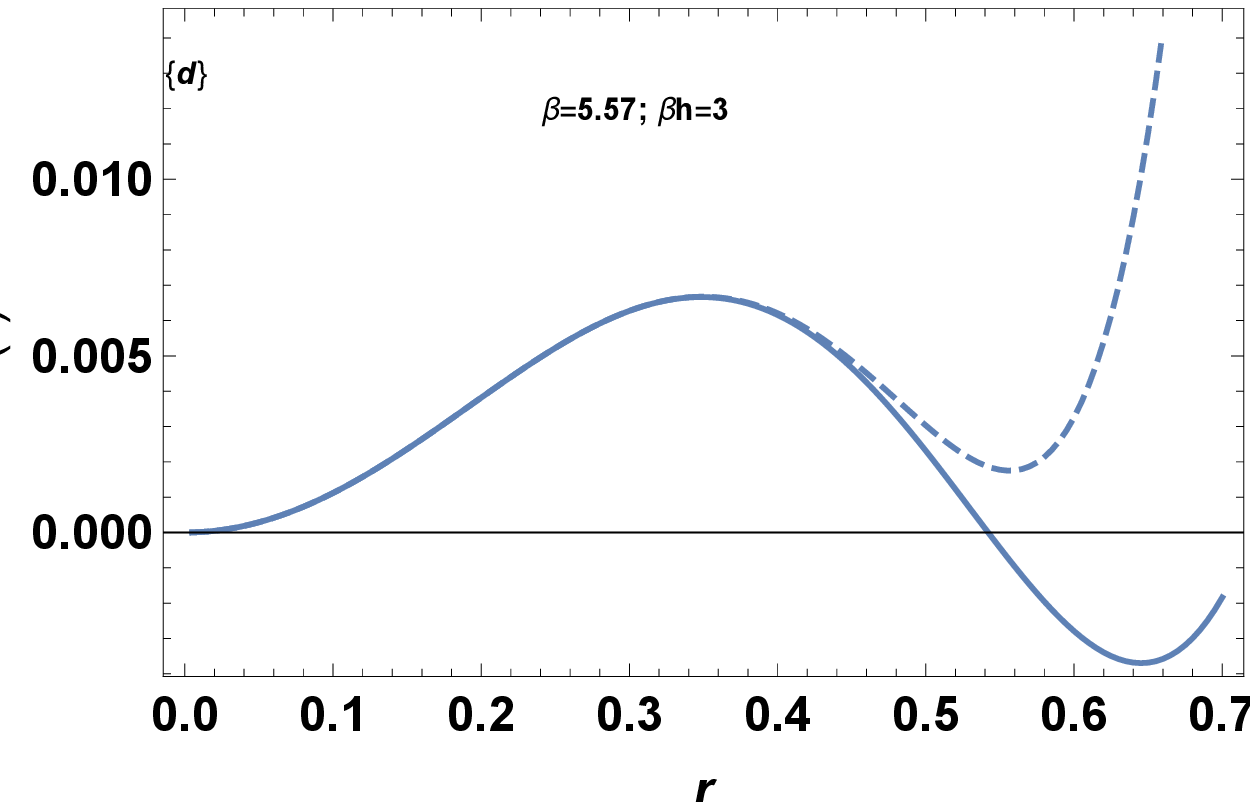}
     \hfill
             \caption{The full rate function $I(r)$ (solid line) and the Landau functional (dotted line) obtained by truncating at the tenth order are plotted as a function of $r$ for two values of $\beta h$, for the uniform distribution. For $\beta h = 1.5$ the two functions are indistinguishable in the plot. There is a continuous transition at $\beta \simeq 3.1026$, so that the minimum of $I(r)$ at $r=0$ for $\beta = 3.1025$ (a) evolves continuously to a nonzero value of $r$ at $\beta = 3.1027$ (b). By contrast, for $\beta h = 3.0$, there is a first order transition at $\beta \simeq 5.56$, so that the minimum of $I(r)$ at $r=0$ for $\beta = 5.55$ (c) jumps to $r=0.645$ at $\beta = 5.57$ (d).}
\label{ratefp}             
\end{figure}

We call the truncated functional in Eq. \ref{lc6} the Landau functional $L(r)$. Since the expression of $c_6$ is long, we have not displayed it; rather we have plotted $c_6$ along with $c_4$ in Fig. \ref{c4}. We observe that as $\beta h$ is increased, $c_4$ changes sign from positive to negative values when
$\beta h$ crosses $1.833$. Similarly, the coefficient $c_6$ also changes sign and is slightly negative for $0<\beta h<1.038$, then positive for $1.038<\beta h<3.318$ and then negative for larger values of $\beta h$. Hence the truncated functional in Eq. \ref{lc6} can only be used to get the phase boundary for $\beta h <1.833$. The plots of the full function $I(r)$ and the Landau function $L(r)$ are shown in 
Fig. \ref{ratefp} (a) and (b) for $\beta h=1.5$ and are seen to be practically indistinguishable for small $r$. 

Hence for $\beta h< 1.833$, the model has a continuous transition from a paramagnet to an $XY$ ferromagnet. For a given strength of $h$, the critical value of $\beta_c$ is obtained by setting $c_2=0$, which yields:
\begin{equation}
1-\frac{\beta_c}{2}+\frac{\beta_c I_1(\beta_c h)^2}{2 I_0(\beta_c h)^2}=0
\label{rfcpu}
\end{equation}
We will show later that this equation of the critical line is valid for any quadriperiodic distribution $p(\alpha)$.

For the uniform distribution, if $\beta h > 1.833$ we need to study the full rate function numerically. We find a first order transition between the paramagnetic and the ferromagnetic states (see Fig. \ref{ratefp} (c) and (d)). The abrupt change in the location of the absolute minimum of the rate function confirms the  first order transition. The function $L(r)$ obtained by expanding $I(r)$ till tenth order in $r$ also predicts a first order transition but with a small error in the location of the transition. 

Hence the model displays regions of second order and first order transitions in the $T-h$ plane. The second order transition line ends at $\beta h =1.832$ and $\beta=3.614$. At this point $c_4=0$ and hence according to Landau theory, this is a tricritical point where the exponents change from mean field Ising values to the mean field tricritical universality class. Beyond this, the model exhibits a line of first order transitions. In Fig. \ref{pdrf} we 
give the complete phase diagram in the $T-h$ plane. The tricritical point at $\beta h=1.832$ and $\beta=3.614$ ($T=0.277$ and $h=0.507$) is denoted by the black circle in the phase diagram  in Fig. \ref{pdrf}.


The region below the transition lines in the $T-h$ plane is actually a phase co-existence surface, where multiple phases can coexist. Each such phase can be stabilized by adding a guiding field conjugate to the magnetization. For an isotropic distribution of random fields, there is evidently an infinite number of directions for the guiding field to point along, and hence an infinity of possible phases. For the distribution with cubic symmetry discussed below, there are four relevant ordered phases.

We note that the locus of critical points given by Eq. \ref{rfcpu} is the same as obtained in the dense limit of regular random graphs using the belief propagation method \cite{lupo}. In \cite{lupo}, this equation was assumed to be the phase-boundary in the full $T-h$ plane, terminating at  $(T,h)=(0,0.5)$. In actuality, the equation gives the phase boundary only for $\beta h<1.832$, as revealed by our study of the full rate function \cite{sumedhaetal}, corroborated by \cite{lupoc}.

\subsection{Cubic Distribution}
\begin{figure}
     \centering
         \includegraphics[width=0.45\textwidth]{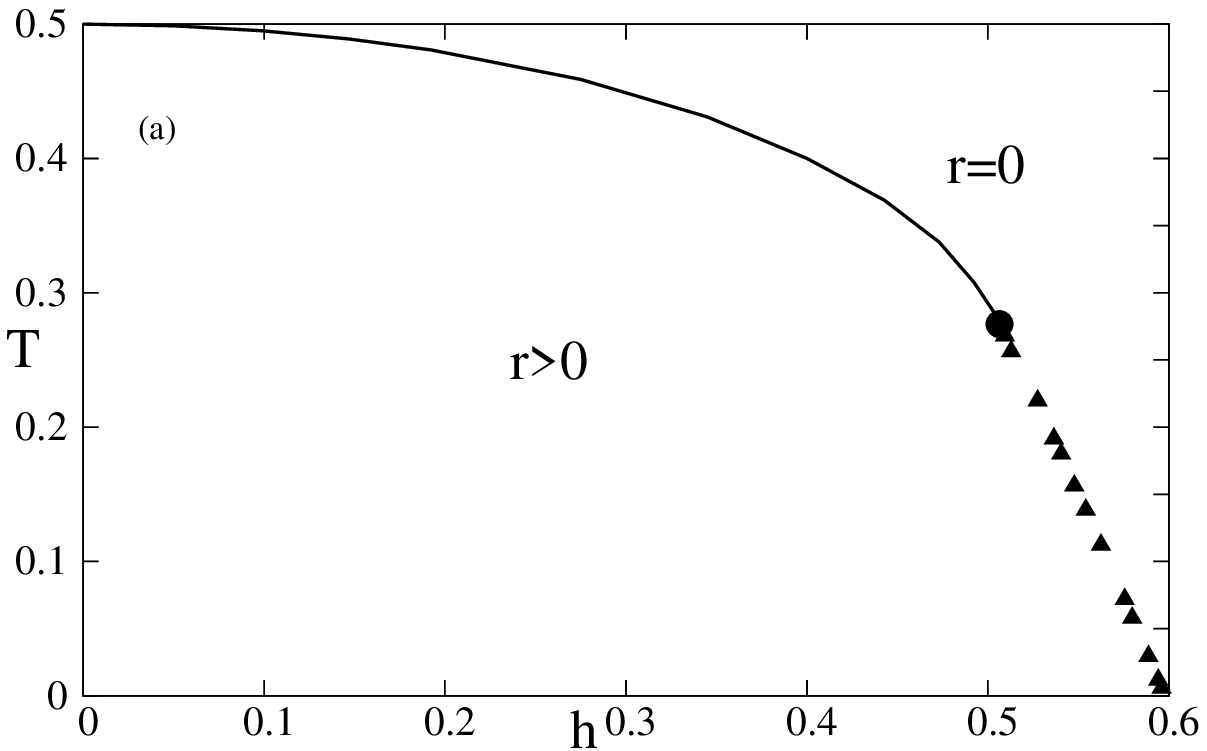}
         \includegraphics[width=0.45\textwidth]{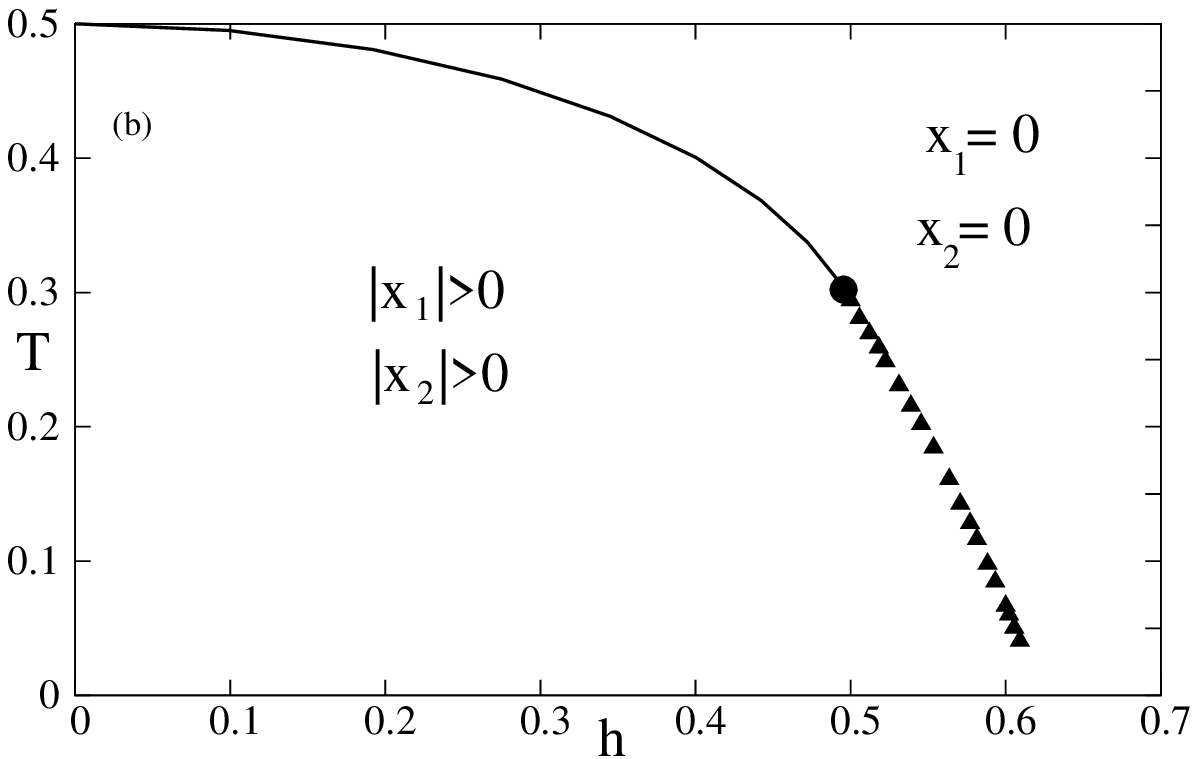}
     \hfill
             \caption{Phase diagrams, with (a) a uniform distribution (b) a distribution with cubic symmetry. Solid lines depict the loci of continuous transitions, while triangular points depict loci of first order transitions. The black dot is the multicritical point.}
        \label{pdrf}
\end{figure}
We now study a symmetric cubic distribution in which the random field points along one of the $4$ directions( $\pm \hat{i}$,$\pm \hat{j}$), i.e the angle $\alpha$ is chosen from the following distribution:
\begin{equation}
p(\alpha)=\frac{1}{4}\left(\delta(\alpha-0)+\delta(\alpha-\pi/2)+\delta(\alpha-\pi)+\delta(\alpha-3 \pi/2)\right)
\end{equation}
In this case the rate function becomes:
\begin{eqnarray}
\label{cubicrf}
I(x_1,x_2)&=& \frac{\beta}{2} (x_1^2+x_2^2)+\log I_0(\beta h)-\frac{1}{4} \sum_{s = \pm 1}\\\nonumber
&&(\log I_0(\sqrt{\beta^2 h^2+\beta^2(x_1^2+x_2^2)+2 s \beta^2 h x_1})+\\\nonumber
&&\log I_0(\sqrt{\beta^2 h^2+\beta^2(x_1^2+x_2^2)+2 s \beta^2 h x_2}))
\end{eqnarray}
Note that in general $I(x_1,x_2)$ is a symmetric function of the two variables $x_1$ and $x_2$, and not a single combination as in the uniform case. On expanding till fourth order, we obtain the Landau functional,
\begin{equation}
L(x_1,x_2)= c_2(x_1^2+x_2^2)+\frac{c_4}{h^4} (x_1^4+x_2^4)+2 \frac{c_{22}}{h^4} x_1^2 x_2^2
\label{lc}
\end{equation}

The coefficient $c_2$ is given by Eq. \ref{c2u}, while
\begin{eqnarray}
c_4&=&\frac{a}{24 I_0^4}(a (a^2-3) I_0^4+(6-4 a^2) I_0^3 I_1+a(7-4 a^2) I_0^2 I_1^2\\\nonumber
&+&6 a^2 I_0 I_1^3+3 a^3 I_1^4)
\label{c4c}
\end{eqnarray}
\begin{eqnarray}
c_{22}&=&\frac{1}{4 I_0^3} (a^2 I_0^3-2 a I_0^2 I_1-a^2 I_0 I_1^2-2 a^3 I_1^3+2 a^2 I_0^2 I_2+2 a^3 I_0 I_1 I_2)
\label{c6c}
\end{eqnarray}
In the above equations, $a=\beta h$ and $I_k$ is the $k^{th}$ modified Bessel function of the first kind with argument $a$. 

Notice that $c_4 \neq c_{22}$ in general. The 
extremum points of the functional defined in Eq. \ref{lc} 
are given by the following two equations:
\begin{eqnarray}
x_1(c_2+2 \frac{c_4}{h^4} x_1^2+2 \frac{c_{22}}{h^4} x_2^2)&=&0\\
x_2(c_2+2 \frac{c_4}{h^4} x_2^2+2 \frac{c_{22}}{h^4} x_1^2)&=&0
\end{eqnarray}
These two equations have four possible solutions:$(0,0)$,$(0,\sqrt{-c_2 h^4/(2 c_4)})$,$(\sqrt{-c_2 h^4/(2 c_4)},0)$ and $(\sqrt{-c_2 h^4/(2 c_4+2 c_{22})},\sqrt{-c_2 h^4/(2 c_4+2 c_{22})})$(one should consider  only the positive roots). If $c_2>0$ then the only stable state is $(0,0)$. For $c_2<0$, if the ratio $c_4^2/c_{22}^2>1$,
the states with $x_1 \neq 0$ and $x_2 \neq 0$ is chosen, whereas for $c_4^2/c_{22}^2<1$, states $(0,\sqrt{-c_2 h^4/(2 c_4)})$ and $(\sqrt{-c_2 h^4/(2 c_4)},0)$ will be stable.
\begin{figure}
\centering
\includegraphics[scale=0.7]{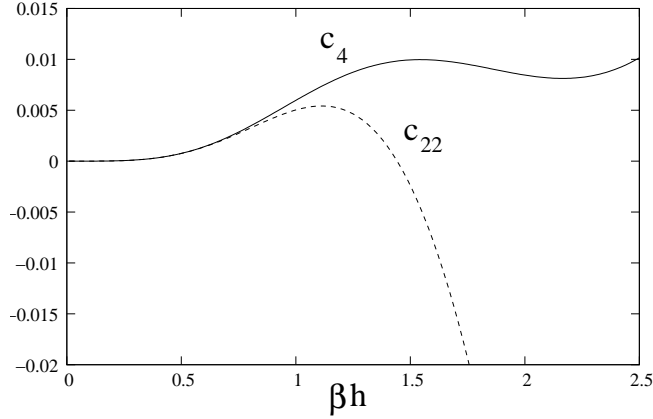}
\caption{The solid line shows the coefficient $c_4$ while the 
dotted line shows the coefficient $c_{22}$ in the expansion of $I(x_1,x_2)$ for the cubic distribution of the random field orientation. Note that the coefficient $c_{22}$ changes sign at $\beta h=1.442$.}
\label{c4c22}
\end{figure}

\begin{figure}
     \centering
         \includegraphics[width=0.45\textwidth]{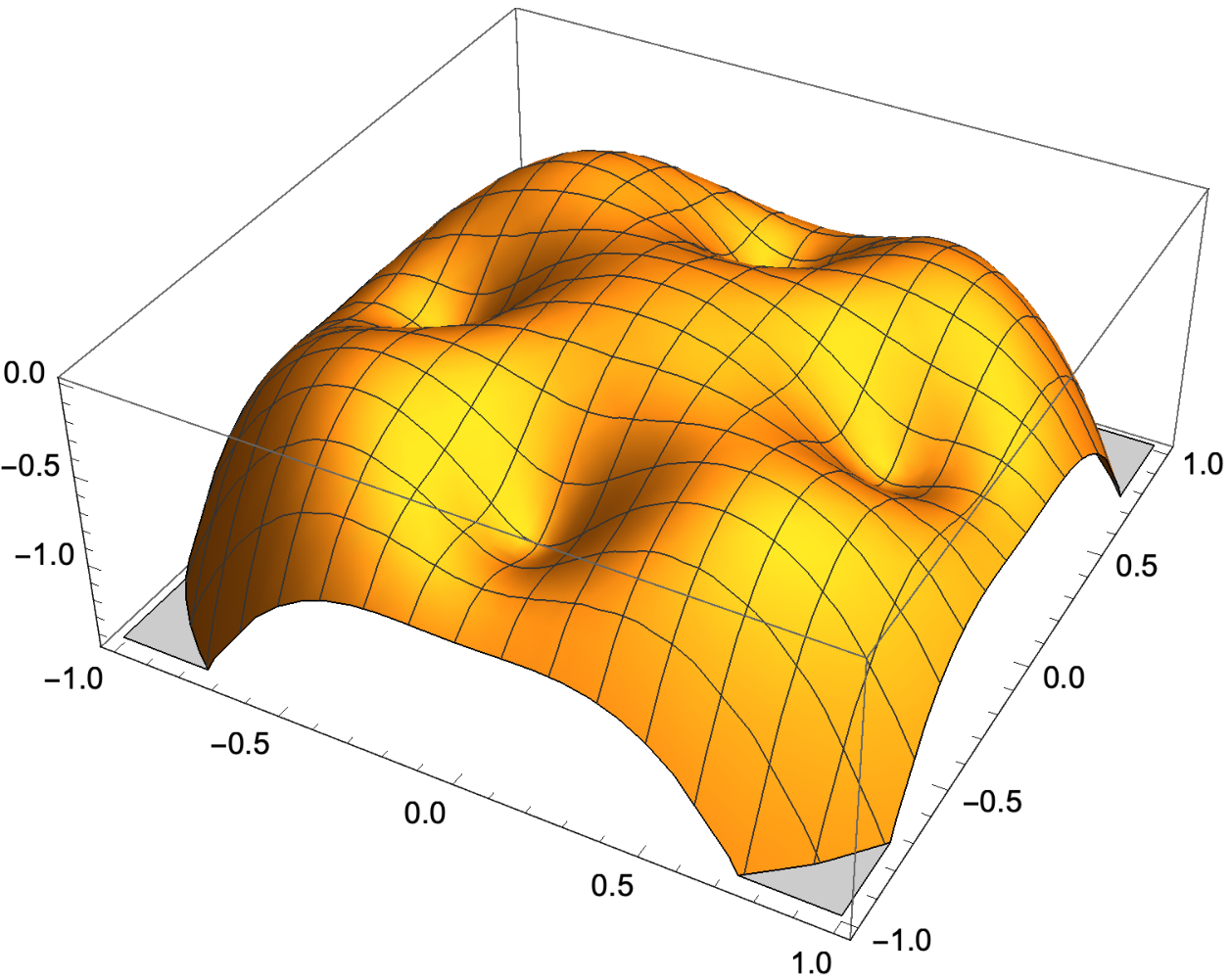}
     \hfill
         \includegraphics[width=0.45\textwidth]{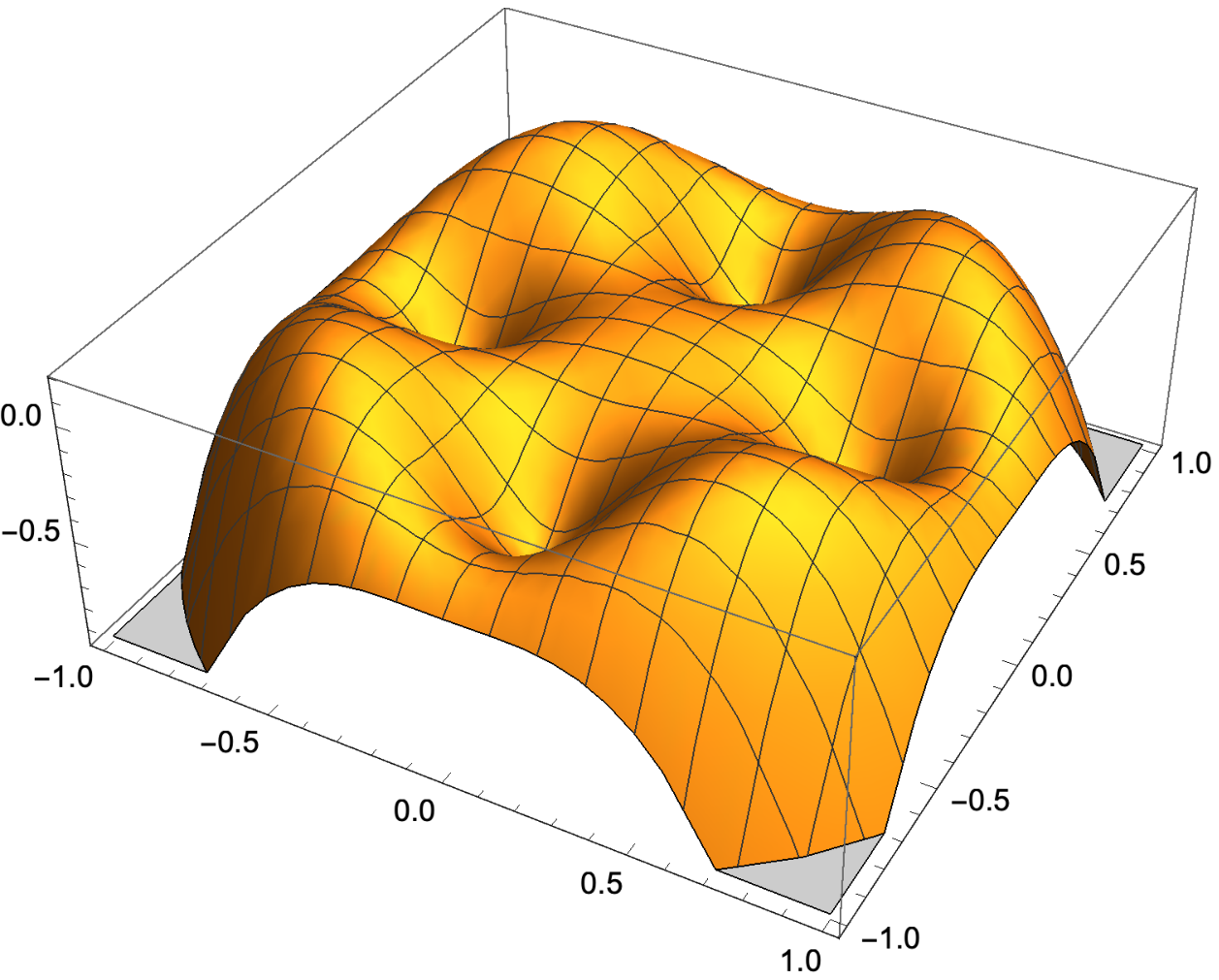}
     \hfill
             \caption{The negative of $I(x_1,x_2)$ is plotted as a function of $x_1$ and $x_2$ for $\beta h =10$ and $h=0.58$ (left) and $h=0.62$ (right). The local maxima in the figure are local minima of $I(x_1,x_2)$. For $h=0.58$, the global mimimum is at $(0,0)$ while for $h=0.62$ there are four coexisting global minima of $I(x_1,x_2)$.}
        \label{3dI}
\end{figure}


Note that once the combination $(c_4+c_{22})$ becomes negative, we cannot use Eq. \ref{lc} to determine the locus of transitions, as the transition becomes first order. 
As $\beta h$ is varied, we find that the coefficient $c_4$ is always positive. We have checked this numerically, as shown in Fig. \ref{c4c22}; also as $a = \beta h \rightarrow \infty$, we have $I_1/I_0 \rightarrow 1$, and $c_4 = \frac{a (3+2 a+a^2)}{12}$, which is positive. On the other hand, the coefficient of $c_{22}$ changes sign from positive to negative at $a=\beta h =1.442$ as shown in Fig.  \ref{c4c22}. The combination $(c_4 +c_{22}) \ge 0$ for $\beta h \le (\beta h)_c =1.631$, and becomes negative for larger values of $\beta h$. By studying the entire rate function, we confirm the first order transition for 
$\beta h > (\beta h)_c$. For $\beta h < (\beta h)_c$, the critical locus is found by equating $c_2$ to zero, which gives Eq. \ref{rfcpu}.

For $\beta h =(\beta h)_c$, there is a continuous transition at $h_c=0.495$ and $\beta =3.294$($T=0.304$) which falls in the tricritical Ising universality class. The ratio $(c_4/c_{22})^2$ exceeds unity for $\beta h < (\beta h)_c$ and becomes less than one beyond that. Hence, we find that the transition line given by Eq. \ref{rfcpu} separates the $(x_1, x_2)=(0,0)$ state from four degenerate states with $(x_1, x_2)$ given by $\pm (\sqrt{-c_2/(2 c_4+2 c_{22}) },\pm \sqrt{-c_2/(2 c_4+2 c_{22}) } )$.

For $\beta h > (\beta h)_c$, we obtain the transition point by graphically studying the full rate function given in Eq. \ref{cubicrf}. We find a first order transition from the 
state $(0,0)$ to state with four degenerate states as 
mentioned above. These are illustrated in Fig. \ref{3dI} for $\beta h =10$. For better visualisation, we have plotted the negative of the rate function, so that the maxima in the figure are actually minima of $I(x_1,x_2)$. The system undergoes a first order transition at $h=0.604$ in this case. In the figure we have plotted the function for $h=0.58$ and $h=0.62$. In both cases, one can see five local minima of $I(x_1,x_2)$. For $h=0.58$, the global minimum is at $x_1=x_2=0$, while for $h = 0.62$, $I(x_1,x_2)$ is minimum at the four other degenerate states. The phase diagram for the distribution with cubic symmetry is shown (Fig. \ref{pdrf}). 
\subsection{Quadriperiodic Distribution}

We show below that the locus of continuous phase transitions is given by a single equation for a family of quadriperiodic distributions, which includes the uniform and cubic distributions as special cases.

Expanding Eq. \ref{grf} to quadratic order, we obtain the term
\begin{align}
 & \frac{\beta (x_1^2+x_2^2)}{2} - \int_0^{2 \pi} d \alpha~p(\alpha) \left(\frac{\beta^2}{2} \left(1-\frac{I_1^2}{I_0^2}\right) (x_1^2 \cos^2\alpha+x_2^2 \sin^2\alpha)- \frac{\beta}{2 h} \frac{I_1}{I_0}(x_1^2-x_2^2) \cos 2 \alpha) \right)\nonumber \\
 &= c_2 (x_1^2+x_2^2)+d_2 (x_1^2-x_2^2) \int_0^{2 \pi} d \alpha~p(\alpha) \cos 2 \alpha
 \label{generalc2}
\end{align}
where $c_2$ is the same as in Eq. \ref{c2u} and  $d_2$ is:
\begin{equation}
d_2 = \frac{\beta}{2} \left(\frac{I_1}{I_0 h} - \frac{\beta}{2} \left(1-\frac{I_1^2}{I_0^2}\right)\right)
\end{equation}
A sufficient, though not necessary, condition for $d_2$ to vanish is that $p(\alpha)$ be quadriperiodic, namely
\begin{equation}
p(\alpha) = p(\alpha+\pi/2)
\end{equation}
We conclude that the line of continuous transitions is given by the same equation for all quasiperiodic distributions. The second order locus terminates at a tricirtical point which is different for different distributions. Also as seen in the case of cubic and uniform distributions, the nature of the ordered state is also different for different symmetric distributions.

This feature will also be shared by even more general distributions for which the disorder average of $\cos 2 \alpha$ is zero. For example, let us consider an asymmetric distribution:\begin{equation}
p(\alpha)=p_1 \delta(\alpha-0)+p_2 \delta(\alpha-\pi/2)+p_3 \delta(\alpha-\pi)+p_4 \delta(\alpha-3 \pi/2)
\end{equation}
where $p_1+p_2+p_3+p_4=1$. In this case the second order term in the expansion of $I(x_1,x_2)$ is
\begin{equation}
c_2 (x_1^2+x_2^2)+d_2 (p_1+p_3-p_2-p_4) (x_1^2 - x_2^2)
\end{equation}
For $p_1+p_3=p_2+p_4$, the locus of continuous transitions (if it exists) is given by $c_2=0$.

\section{Thermodynamic quantities}
\label{sec3}
\begin{figure}
  \centering
         \includegraphics[width=0.33\textwidth]{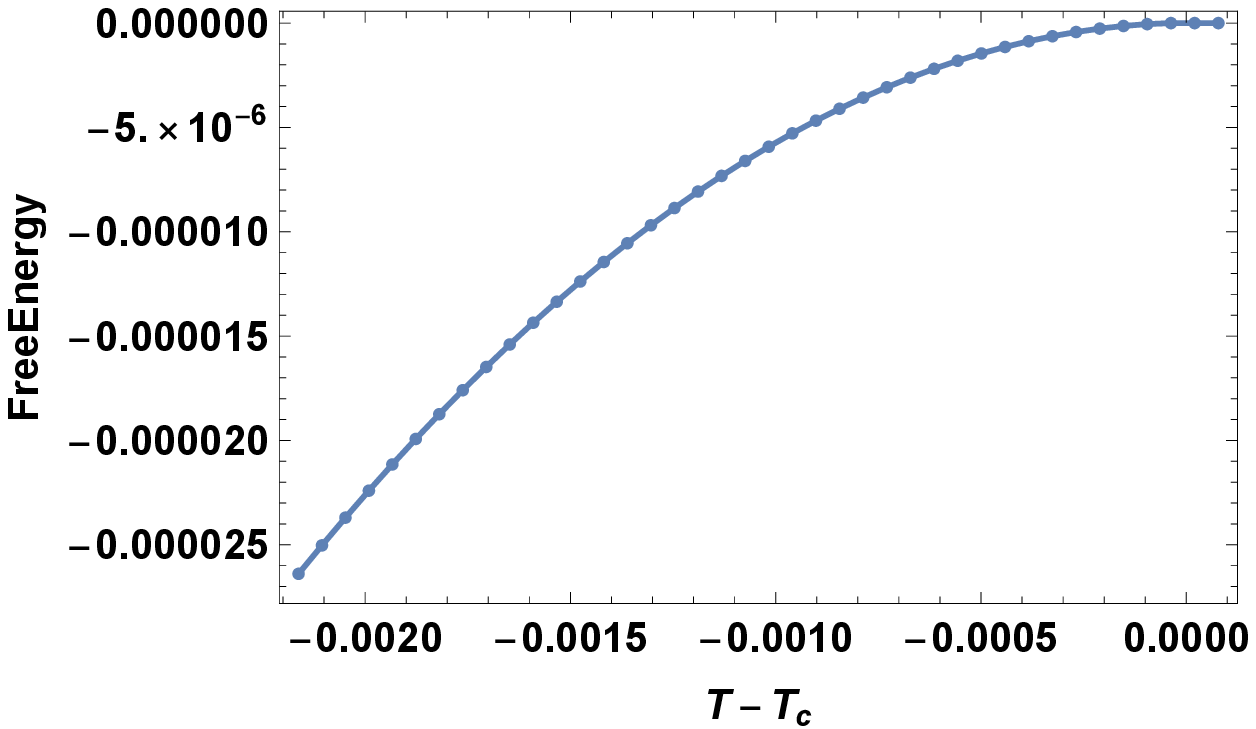}
         \includegraphics[width=0.3\textwidth]{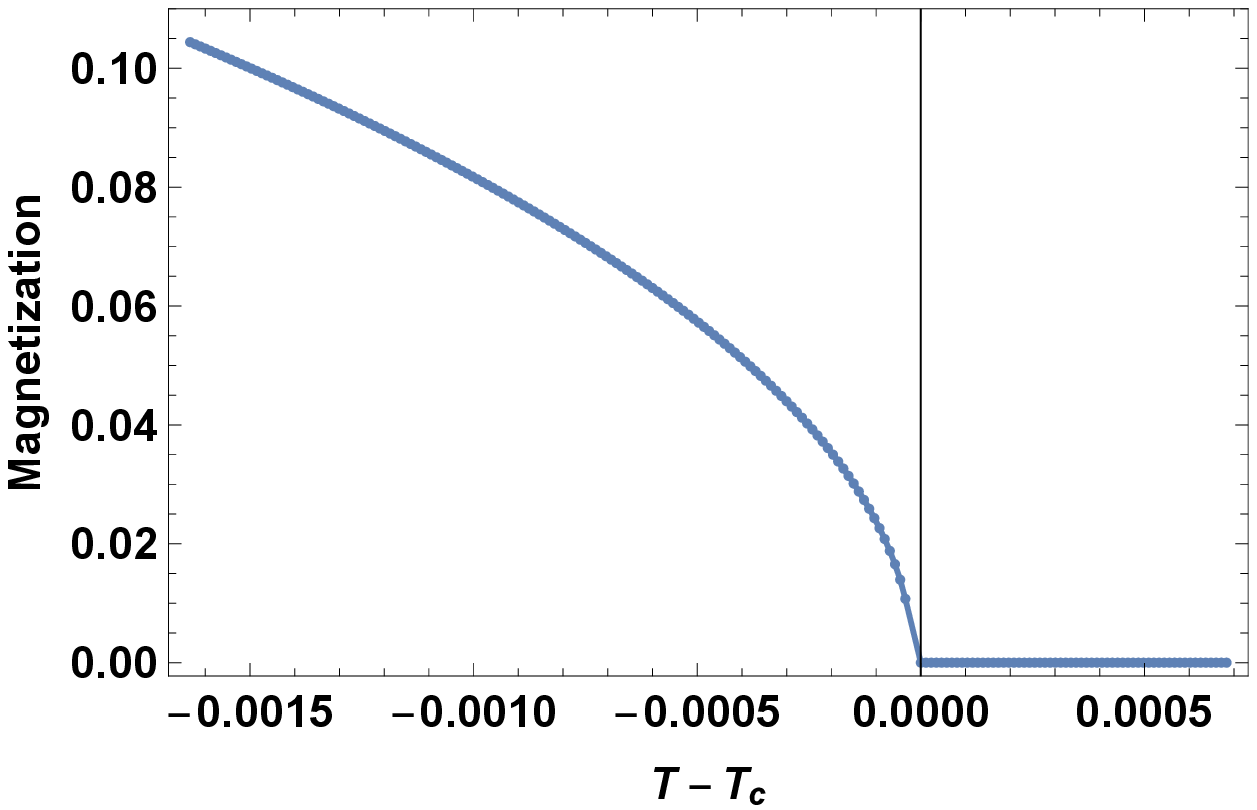}
     \includegraphics[width=0.3\textwidth]{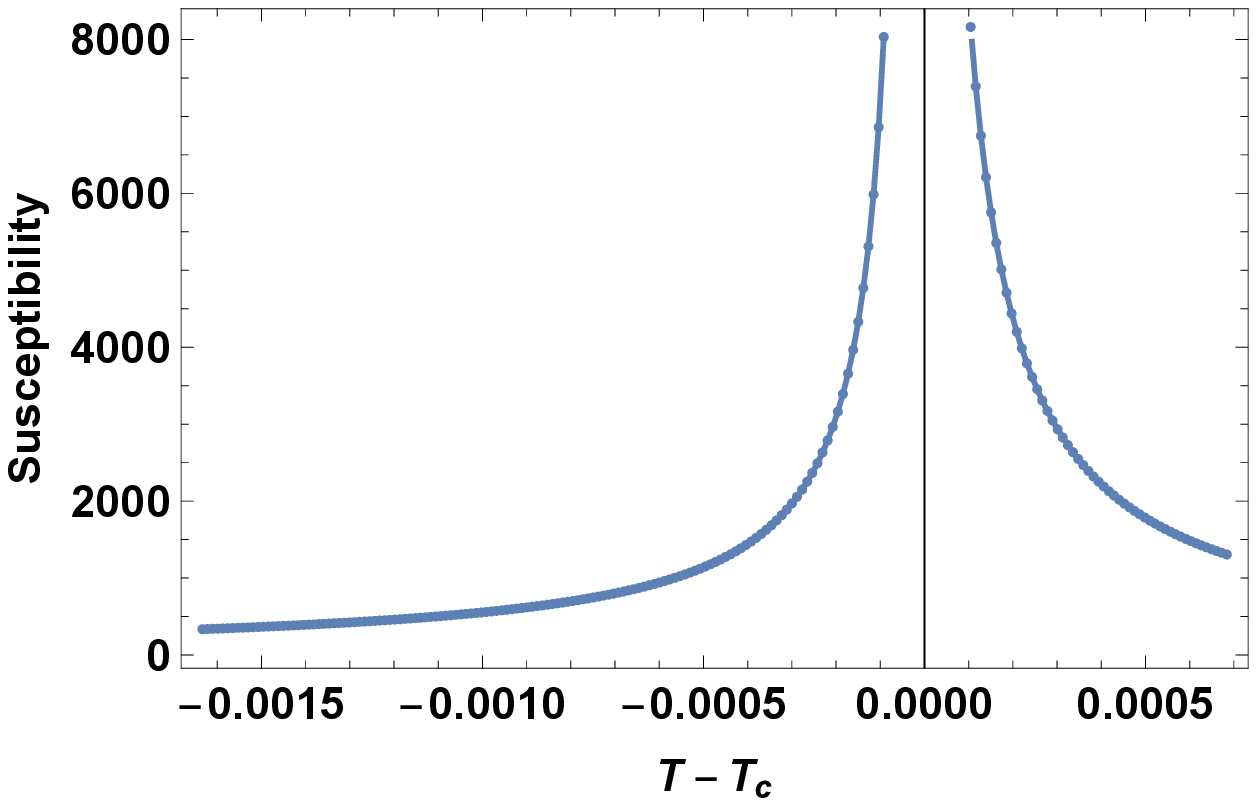}
     \hfill
             \caption{Free energy, magnetization and magnetic susceptibility at $h=0.47$ near the continuous transition at  $T_c=0.3400$ for the cubic distribution of disorder. The free energy is fitted well by the function $-5.68 |T-T_c|^2$ for $T<T_c$, implying that the specific heat exponent $\alpha =0$. The magnetization is fitted by $2.66 (T_c-T)^{0.5}$ for $T<T_c$, consistent with $\beta =0.5$. The transition is continuous and lies in the mean field Ising universality class. The susceptibility diverges as $|T-T_c|^{-1}$ from both sides of $T_c$, with amplitude ratio equal to $2$.}
        \label{contc}
\end{figure}
Thermodynamic quantities become anamolous as the phase boundary is approached. Their behaviour depends on whether the approach is to (i) the line of continuous transition, (ii) the tricritical point, or (iii) the first order line. This is illustrated below by examining the free energy, magnetization and susceptibility of the cubic model, in regions (i), (ii) and (iii).

In the disordered phase ($T>T_c$), all these quantities behave in a similar fashion in all three regions. The magnetization is zero, and consequently the free energy vanishes as well. The susceptibility ($\chi$) can be found as follows. Add a term $-\beta H x_1$ to $I(x_1,x_2)$ in Eq. \ref{ratefp} and minimize w.r.t $x_1$ to find the self-consistent equation for $x_1$, and then find $\chi_1 \equiv \partial x_1/\partial H$. In the limit $H \rightarrow 0$, we obtain,
\begin{equation}
\chi_1 = \frac{\beta}{\frac{\partial^2 f(x_1,x_2)}{\partial x_1^2}-\beta}
\end{equation}
where  $f(x_1,x_2) =\int_0^{2 \pi} d \alpha~p(\alpha) \log I_0(\beta \sqrt{h^2+(x_1^2+x_2^2)+2 h (x_1 \cos \alpha+x_2 \sin \alpha)})$.

For the cubic distribution, we find
\begin{equation}
\chi_1 = \frac{2 T I_0^2}{(-1+2 T) I_0^2+I_1^2}
\end{equation}
On noting that $I_1^2 = (1-2 T_c) I_0^2$, where $T_c$ is the critical temperature along the critical curve, we see that $\chi$ reduces to the Curie-Weiss form
\begin{equation}
\chi_1= \frac{1}{T-T_c}
\end{equation}
This holds in the $T>T_c$ regions of (i),(ii) and (iii).

In the ordered phase ($T<T_c$), there are some marked differences of behaviour in the regions (i),(ii) and (iii), These are brought out below.

Figure \ref{contc} shows the behaviour of these quantities as a function of temperature along the locus $h=0.47$, which lies in the region (i), since $(c_4+c_{22}) >0$, the system undergoes a normal second order transition, governed by Landau theory. Correspondingly (a) the free energy vanishes above $T_c$, and approaches zero quadratically in ($T_c-T$) as $T \rightarrow T_c^{-}$, tentamount to a discontinuity of the specific heat, (b) the magnetization varies as $(T_c-T)^{\beta}$ with $\beta=1/2$ and (c) the susceptibility diverges as $|T_c-T|^{-\gamma}$ with $\gamma=1$.

\begin{figure}
         \includegraphics[width=0.33\textwidth]{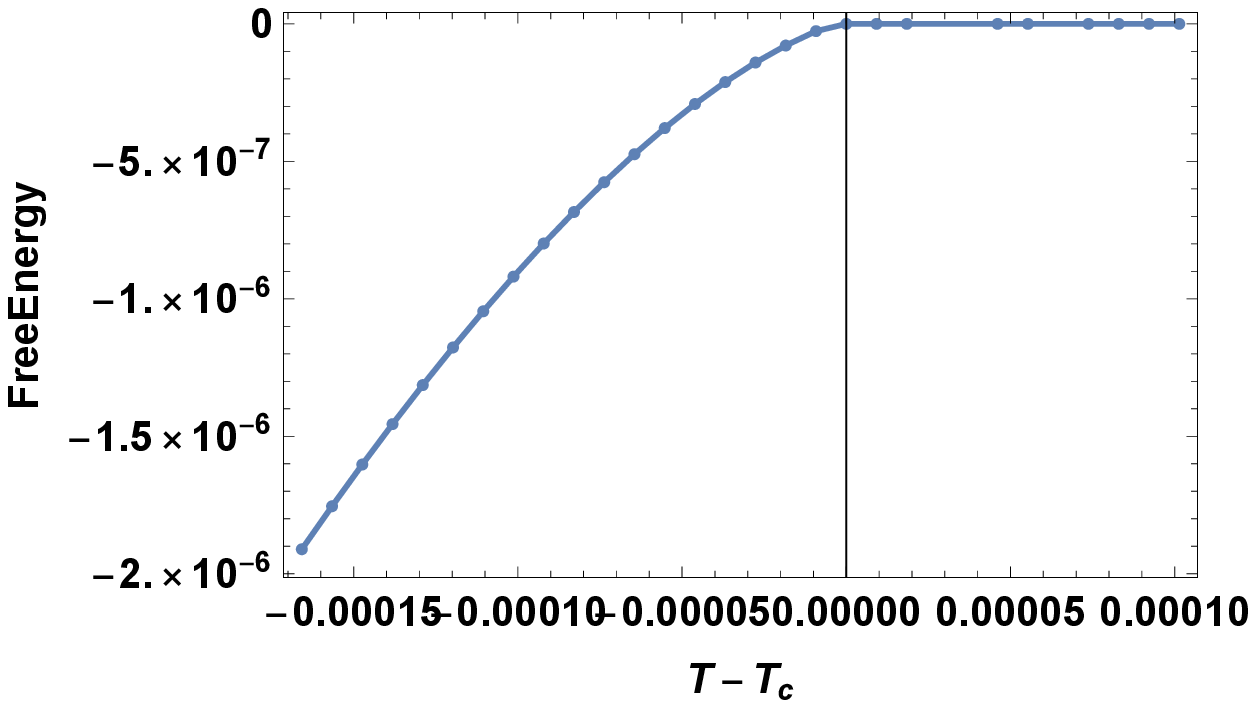}
         \includegraphics[width=0.3\textwidth]{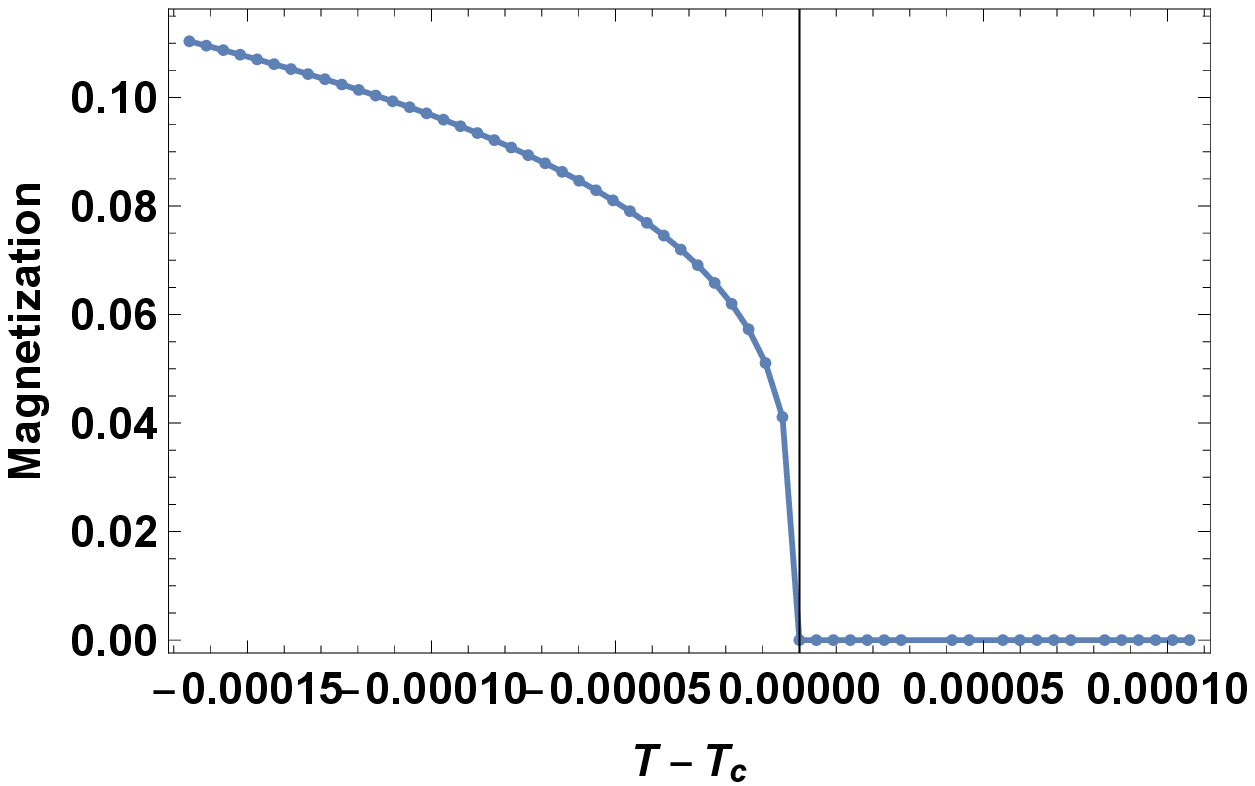}
     \includegraphics[width=0.31\textwidth]{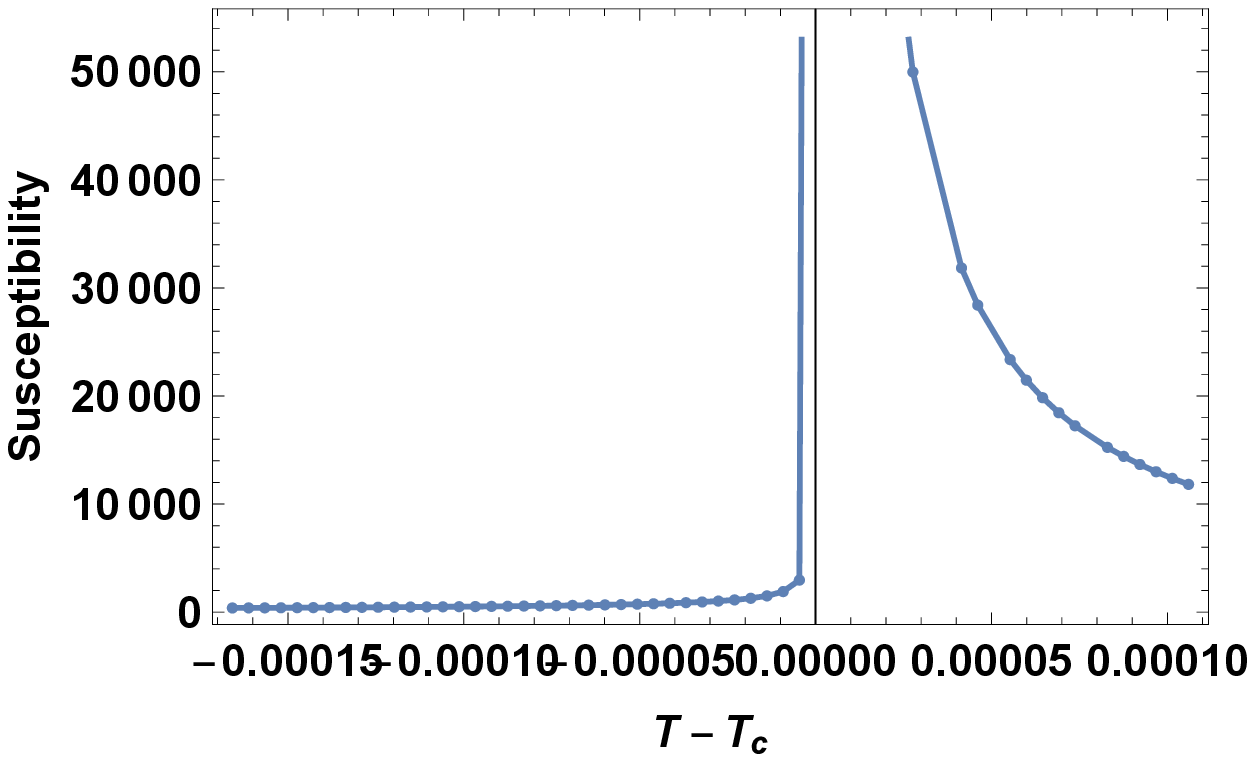}
     \hfill
             \caption{Free energy, magnetization and magnetic susceptibility near the tricritical point ($h_c=0.494984$ ,$T_c=0.303564$) for cubic disorder. The free energy is fitted well by the function $-0.81 |T-T_c|^{1.5}$, which implies that the specific heat exponent $\alpha=0.5$. The magnetization plot is fit by $1.06 (T_c-T)^{0.26}$ for $T<T_c$, consistent with $\beta =0.25$. The susceptibility diverges as $|T-T_c|^{-1}$. The transition is continuous and lies in the mean field tricritical Ising universality class. }
\label{tcpc}
\end{figure}

The locus $h=0.494984$ shown in Fig. \ref{tcpc} describes an approach to the tricritical point $(c_4+c_{22})=0$, corresponding to the region (ii) in which the fourth order term in the Landau theory vanishes. Correspondingly, (a) the free energy vanishes as $(T_c-T)^{3/2}$ as $T \rightarrow T_c^{-1}$, (b) the magnetization follows $(T_c-T)^{\beta}$, with $\beta=1/4$ and (c) the susceptibility diverges as $|T_c-T|^{-\gamma}$, with $\gamma=1$.

\begin{figure}
     \centering
         \includegraphics[width=0.32\textwidth]{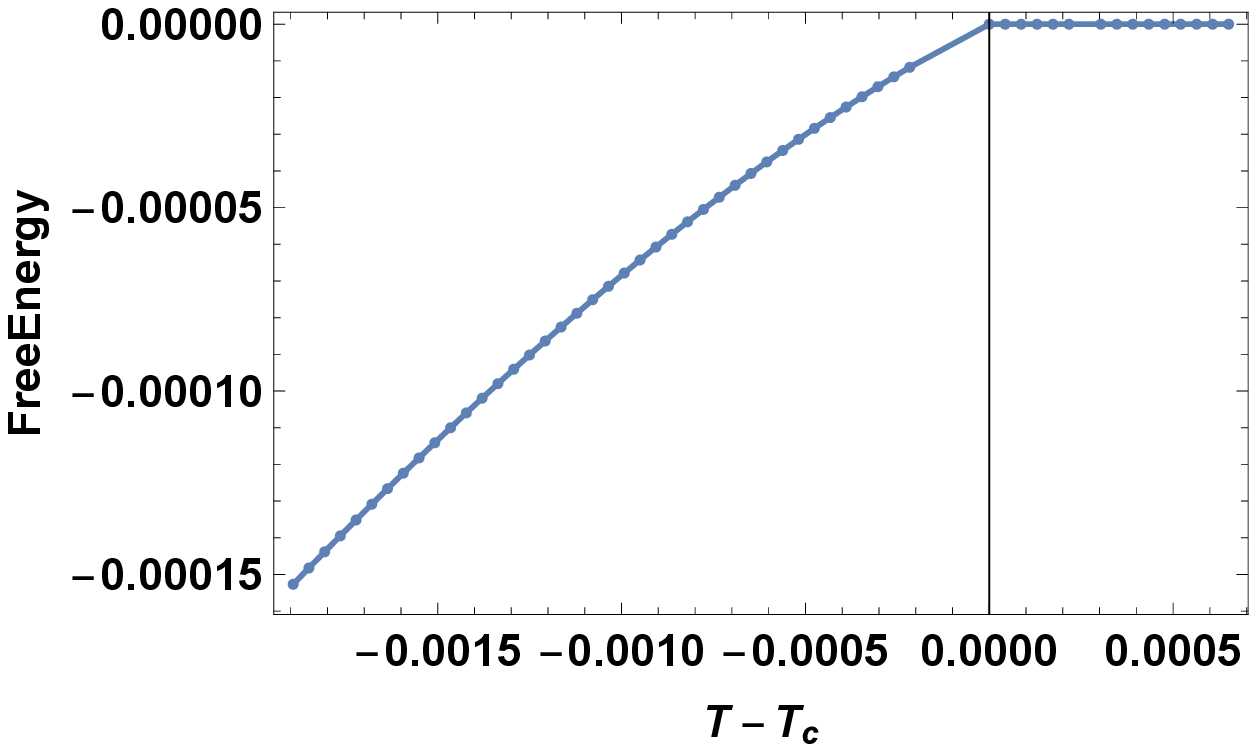}
         \includegraphics[width=0.3\textwidth]{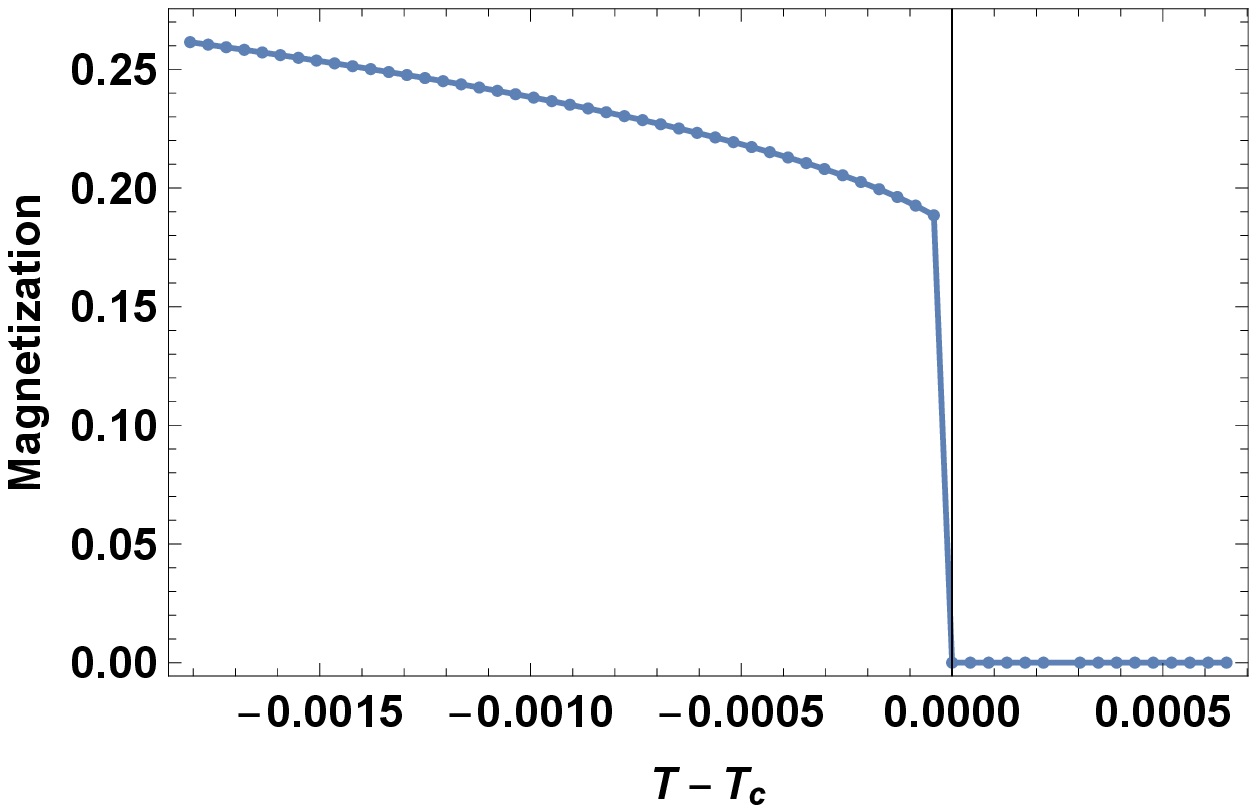}
     \includegraphics[width=0.3\textwidth]{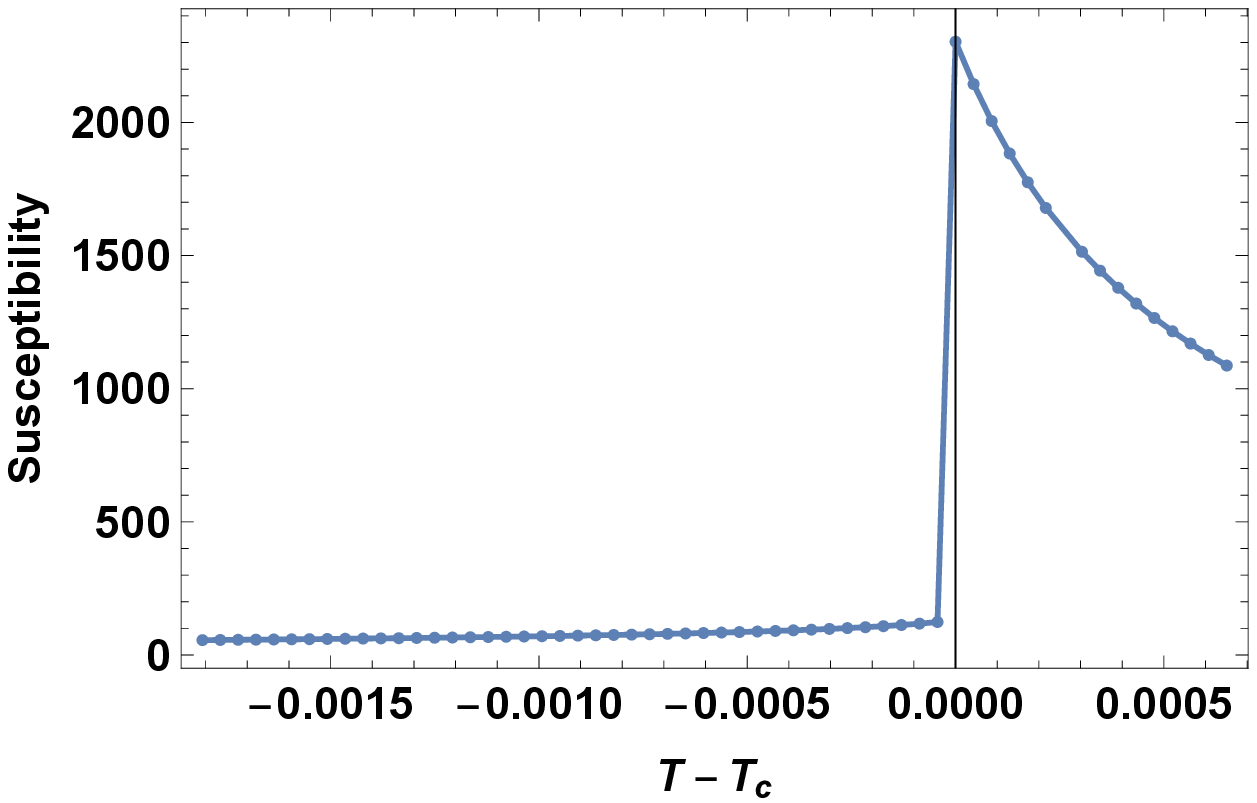}
     \hfill
             \caption{ Free energy, magnetization and magnetic susceptibility at $h=0.5$ near the first order transition at  $T_c=0.2943$ for the cubic disorder. The magnetization and susceptibility show jump discontinuities, while the free energy is linear in the distance from $T_c$ for $T<T_c$, implying a jump in entropy at $T_c$.}
        \label{foc}
\end{figure}

Finally, Fig. \ref{foc} shows the behaviour along the locus $h=0.5$ which lies in the first order region (iii). The free energy has a discontinuous slope, implying that the entropy shows a jump, characterstic of a first order phase transition. Likewise, the magnetization and susceptibility reach finite values as $T \rightarrow T_c^{-}$, implying a jump to values of zero and $1/(T_c-T_0)$ respectively for $T \rightarrow T_c^{+}$, where $T_0$ is the solution of $I_1^2 = (1-2 T_0) I_0^2$. The Curie Weiss form holds, but the divergence is pre-empted by the first order transition.
\section{Low Temperature}
\label{sec4}
\subsection{Ground state energy at zero temperature}
The zero temperature rate function is given by $\phi(x_1,x_2) =\lim _{\beta \rightarrow \infty} \frac{1}{\beta} I(x_1,x_2)$. The disorder averaged ground state energy of the system is the minimum of $\phi(x_1,x_2)$ over $x_1$ and $x_2$.

\begin{figure}
     \centering
         \includegraphics[width=0.45\textwidth]{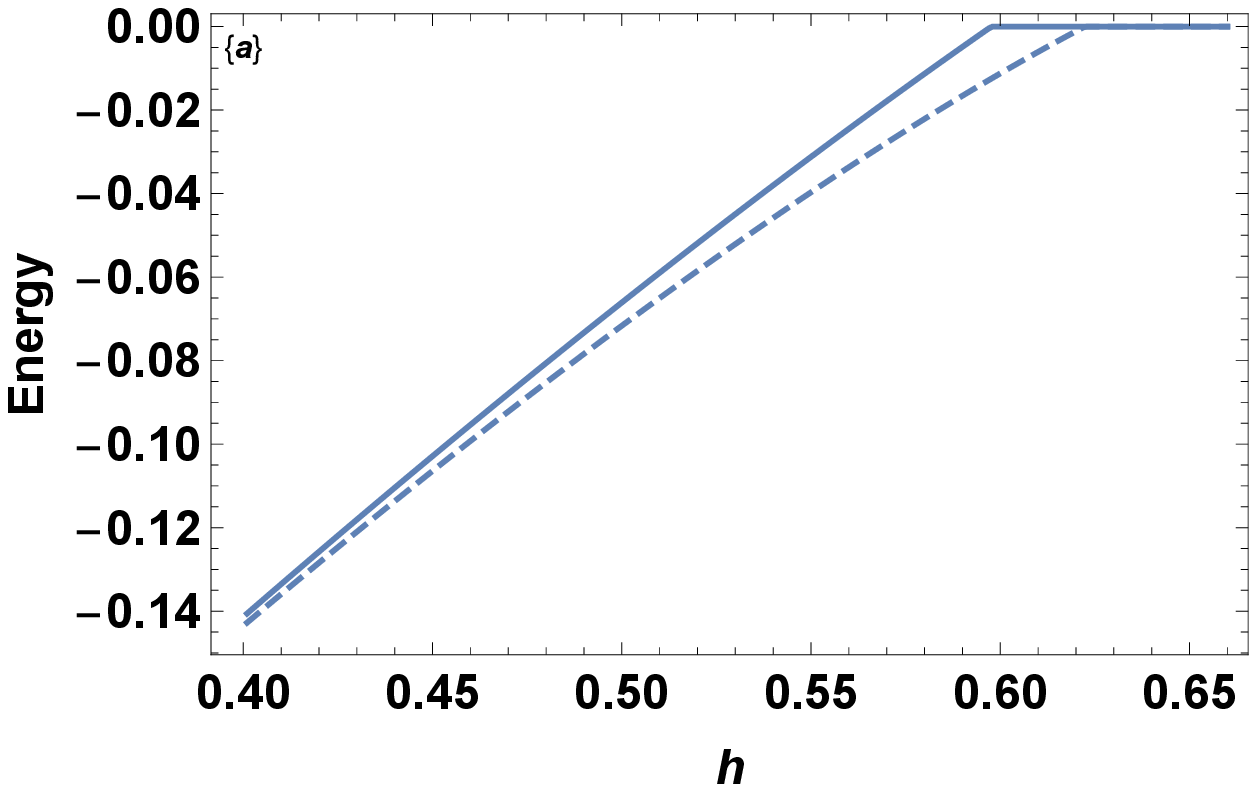}
         \includegraphics[width=0.45\textwidth]{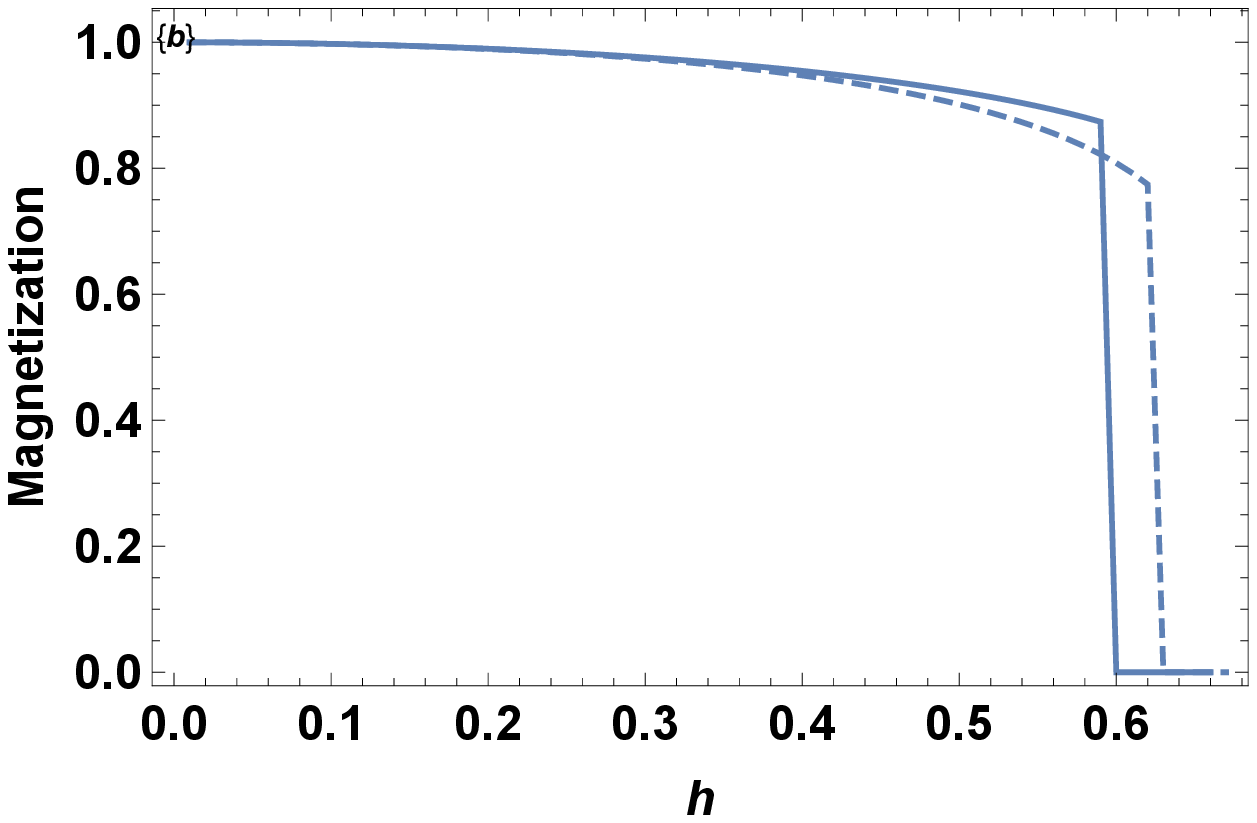}
     \hfill
             \caption{(a) The ground state energy and (b) Magnetization $r$ for uniform (solid line) and cubic (dashed line) distribution for $T=0$. There is a first order transition for both the distributions, but at different values of the random field amplitude.}
        \label{T0}
\end{figure}

For $\beta \rightarrow \infty$, we use the asymptotic form  $I_0(\beta z) \sim \exp (\beta z)$ in Eq. \ref{grf}, and write $x_1=r \cos \theta$ and $x_2=r \sin \theta$, to obtain
\begin{equation}
\phi(r,\theta) = \frac{r^2}{2}+  h - \int_0^{2 \pi} d \alpha~p(\alpha) \sqrt{h^2+r^2+2 h r \cos (\theta -\alpha) }
\end{equation}
For the uniform distribution, we have $p(\alpha)=1/(2 \pi)$, leading to
\begin{align}
\phi(r) &= \frac{r^2}{2} +h -\frac{2 (h+r)}{\pi} E(k)
\end{align}
where $k=\frac{2 \sqrt{h r}}{h+r}$ and $E(k)=\int_0^{\pi/2} d \theta~\sqrt{1-k^2 \sin \theta}$ is the complete Elliptic function of the second kind. The function $\phi(r)$ is greater than $0$ for all values of $r$ for $h>h_c=0.5976\pm 0.0001$, implying that the magnetistaion undergoes a first order jump at $h_c$ as shown in Fig. \ref{T0}.The energy of the system is given by $\phi(0)$ for $h>h_c$ and by $min_r\phi(r)$ for $h<h_c$ (Fig. \ref{T0}).

For the cubic distribution, the function $\phi(r,\theta)$ for a given $r$ has a minimum when $\sin \theta=\cos \theta$. The zero temperature rate function is given by the following equation

\begin{equation}
\phi(r,\pi/4) = \frac{r^2}{2}+h-\frac{1}{2} (\sqrt{h^2+r^2+\sqrt{2} h r}+\sqrt{h^2+r^2-\sqrt{2} h r})
\end{equation}
The function $\phi(r,\pi/4)>0$ for all values of $r$ for $h>h_c=0.6223\pm0.0001$. The magnetization and free energy are plotted 
in Fig. \ref{T0}.

\subsection{Specific heat at low T}
To find the leading low temperature behaviour, we keep the next order term in the asymptotic expansion of $I_0$ and obtain
\begin{align}
I(r,\theta) &= \frac{\beta r^2}{2} + \beta h - \beta \int_0^{2 \pi} d \alpha~p(\alpha) \sqrt{h^2+r^2+2 h r \cos(\theta-\alpha) }\nonumber\\&+\frac{\log(2 \pi \beta)}{2} + \frac{1}{4} \int_0^{2 \pi}d \alpha~p(\alpha) \log (h^2+r^2+2 h r \cos(\theta-\alpha))
\end{align}
Let us confine ourselves to small values of $h$, in which case we can simplify the expression further. We obtain:
\begin{align}
I(r) = \frac{\beta r^2}{2}+\beta h-\beta \sqrt{h^2+r^2}+\frac{1}{2} \log 2 \pi \beta +\frac{1}{4} \log (h^2+r^2)
\label{rflowT}
\end{align}
Setting $dI/dr = 0$ we find
\begin{equation}
 r^2 = r_0^2-T
\end{equation}
where $r_0^2=1-h^2$ is the $T=0$ value for small $h$.  Using this in the expression Eq. \ref{rflowT} for the free energy, and differentiating twice we obtain the specific heat
\begin{equation}
C =\frac{1}{2}-\frac{T}{2} \left(1+\frac{1}{2 r_0^2}\right)
\end{equation}
Note that $C$ approaches a constant as $T\rightarrow 0$, a consequence of low lying excitations, associated with small excursions of the XY spins from their $T=0$ values.

\section{Discussion}
\label{sec5}
Our study of the random-field XY model on the fully connected graph 
using LDT yields the exact free energy, which reveals interesting features of the phase diagram and low-temperature properties for various distributions of the orientation $\alpha$ of the random field. Generically, the phase diagram includes a tricritical point which separates loci of continuous and first order transitions as in Fig. \ref{pdrf}. We identified a broad class of distributions, namely those for which $<cos 2 \alpha> = 0$, all of which share the same locus of continuous transitions Eq. \ref{rfcpu}, although with different tricritical termination points, beyond which the transition is first order. This class includes quadriperiodic distributions, which in turn include the uniform and cubic distributions as special cases.

The equation for the locus of continuous transitions found in \cite{lupo} agrees with our Eq. \ref{rfcpu}, but as we have shown, there is a first order transition beyond the tricritical point and no re-entrance in the phase diagram. It would be interesting to explore whether there is a first order region in the $T-h$ plane in the RFXY model on finitely connected regular random graphs as well.

The nature of the multicritical point which separates the continuous and first order transition loci depends on the distribution of random fields. This is seen by extending the model to include additional uniform fields in different directions. With cubic symmetry in the distribution, four additional critical lines meet at the MCP making a total of five, while 
with an isotropic distribution, the number of possible directions for the ordering field, each
of which induces a new critical line, is infinite.
Turning to low-temperature properties, we obtained an exact expression for the disorder-averaged ground state energy, for an arbitrary distribution of field angles. This allowed us to calculate the value of the field across which the magnetization is discontinuous at $T=0$. We also showed that the low-$T$ specific heat approaches a constant value, indicating the existence of low-lying excited states in this disordered classical spin system.

We conclude with a comment about the models with random crystal field disorder namely the random anisotropy models (RAM), which are relevant to a wide class of disordered magnets. The RAM was solved exactly in the limit of infinite anisotropy on a fully connected graph for uniform distribution of orientations \cite{derrida}. Recently, large deviation theory has been used to solve the model for arbitrary strength of the random anisotropy, for both uniform and bimodal distributions of the orientation \cite{sumedhabarma}. Unlike the RFXY model, for the RAM there is only a continuous transition for both distributions, for all strengths of the random crystal field.

\section{Acknowledgements} 
M.B. acknowledges support under the DAE Homi Bhabha Chair Professorship of the Department of Atomic Energy, India.



\end{document}